\pdfoutput=1

\documentclass[11pt]{article}

\usepackage{acl}
\usepackage{times}
\usepackage{latexsym}
\usepackage{booktabs}
\usepackage[T1]{fontenc}
\usepackage[utf8]{inputenc}
\usepackage{microtype}
\usepackage{inconsolata}
\usepackage{graphicx}
\usepackage{amsmath}
\usepackage{amssymb} 
\usepackage[most]{tcolorbox}
\usepackage{CJKutf8} 
\usepackage{geometry}
\usepackage{makecell}
\usepackage{subcaption}
\usepackage{float} 
\usepackage{fontawesome}

\title{FakeSV-VLM: Taming VLM for Detecting Fake Short-Video News via Progressive Mixture-Of-Experts Adapter}

\author{
Junxi Wang, Yaxiong Wang\textsuperscript{†}, Lechao Cheng, Zhun Zhong\textsuperscript{†} \\
Hefei University of Technology, China \\
\texttt{Junxiwang182@gmail.com}\\
\texttt{wangyx@hfut.edu.cn}, \texttt{chenglc@hfut.edu.cn} \\
\texttt{zhunzhong007@gmail.com} \\
}

\begin{document}
\maketitle

\begin{abstract}

We present FakeSV-VLM in this paper, a new VLM-based framework for detecting fake news on short video platforms. Despite significant efforts to combat this issue due to the severe threat that fake news videos pose to public information security, existing methods still fall short in detection accuracy, often due to lack of knowledge to verify the news is real or not. However, large Vision Language Models (VLMs) have absorbed extensive real-world knowledge from massive multimodal datasets. Motivated by this, we adapt advanced VLMs for fake news detection in short videos.  Upon close examination of news samples, we observe that short video samples can be categorized into four distinct scenarios: both video and text are real (for real samples), or both are fake, or either the video or text is fake (for fake samples). Inspired by this insight, we design four experts tailored to handle each scenario and integrate them into VLM via Mixture of Experts. Specifically, we develop the Progressive MoE Adapter (PMOE) module where detection experts first provide an initial analysis, followed by attribution experts for a comprehensive diagnosis, leading to a robust decision. Additionally, we also note the fake news videos often show inconsistency between two modalities. Consequently, we further design the Alignment-driven Event Checking (ADEC) module, which perceives the fake news by capturing the inconsistency between different modalities. Extensive experiments on two benchmark datasets, FakeSV and FakeTT, verify the superiority of our model. It significantly outperforms current state-of-the-art models by +3.32\% and +5.02\%, establishing a new benchmark in the field. Our code is available at https://github.com/Celina-love-sweet/FakeSV-VLM.

\end{abstract}

\section{Introduction}

With the rapid development of short video platforms, the spread of fake news videos on these platforms has posed significant risks to society, covering key areas such as politics \cite{r5},  economics \cite{r6}, and more. According to relevant statistics, over 500 hours of videos are uploaded to YouTube every minute \cite{r4}. Relying entirely on manual verification of the authenticity of news videos would consume a huge amount of human resources. To address this challenge, some methods for detecting fake news videos have emerged in the early stages \cite{r1,r2,r3}. However, with the continuous evolution of video editing and synthesis technologies, manipulating and fabricating news videos has become increasingly easier \cite{r7}. Moreover, the ongoing development of large language models further enhances the ability to fabricate fake news \cite{r21,r22}, making detection even more difficult and posing an unprecedented threat to public information security.

To address this challenge, several methods have been proposed. FakingRecipe~\cite{r9} predicts video authenticity by fusing visual, textual, and audio features, while NEED~\cite{r23} leverages graph neural networks to analyze relationships among videos from the same event. Although effective, these approaches often rely heavily on cross-modal fusion and are constrained by limited training data, lacking the ability to fully utilize real-world knowledge. Some methods incorporate knowledge from Vision Language Models (VLMs)—for instance, CA-FVD~\cite{r11} detects modality alignment using VLMs, and ExMRD~\cite{r10} distills knowledge through designed reasoning chains—but they typically depend on VLMs to pre-generate text, limiting scalability and practical deployment.

To address the aforementioned issues, we propose directly using VLMs to predict fake news videos. Pre-trained VLMs contain real-world knowledge, which can be fully utilized during the detection process to infer the authenticity of news videos. However, since VLMs are not specifically designed for the task of predicting fake news videos, applying them directly for detection often does not yield ideal results. Based on our observations, news videos can be classified into two broad categories: real and fake. At a more fine-grained level, fake news videos can be subdivided further into content forgery, description forgery, and full forgery of both. 

To enhance the capability of VLMs in modeling hierarchical forgery patterns in fake news videos, we propose the Progressive Mixture-Of-Experts Adapter (PMOE) module. Specifically, a set of learnable Artifact Tokens is introduced to aggregate potential manipulation cues, which are concatenated with multimodal features and fed into the early layers of the LLM. In the intermediate layers, the contextualized Artifact Tokens are extracted for two-stage reasoning: the Detection MoE estimates the overall authenticity of the video, and the Attribution MoE infers the specific manipulation type, such as visual tampering, textual misrepresentation, or cross-modal forgery. This enables fine-grained identification of diverse forgery strategies.

Given the frequent semantic inconsistencies between visual content and textual descriptions in fake news videos, we propose the Alignment-driven Event Checking (ADEC) module to enhance cross-modal event-level consistency verification. ADEC first extracts intermediate textual representations from the LLM and applies average pooling to obtain a compact event-centric semantic embedding. This embedding is then compared with the visual representation using a contrastive learning objective. Supervised by ground-truth labels, the model is encouraged to align cross-modal semantics and detect event-level discrepancies indicative of manipulation.

Based on the above design, we propose the FakeSV-VLM framework to advance multimodal fake news videos detection. The main contributions of this work are summarized as follows:

\noindent 

$\bullet$ We integrate VLMs into the training process and directly apply them to fake news videos detection tasks, driving the development of this field. 
    
$\bullet$ We design the PMOE and ADEC modules to enable robust manipulation reasoning and event-level cross-modal consistency checking, thereby effectively enhancing the VLM's capacity for fake news videos detection.
    
$\bullet$ The FakeSV-VLM framework is end-to-end, highly generalizable, and achieves remarkable results, far surpassing the existing SOTA methods.

\section{Methodology}

\begin{figure*}
  \centering
  \includegraphics[width=\textwidth]{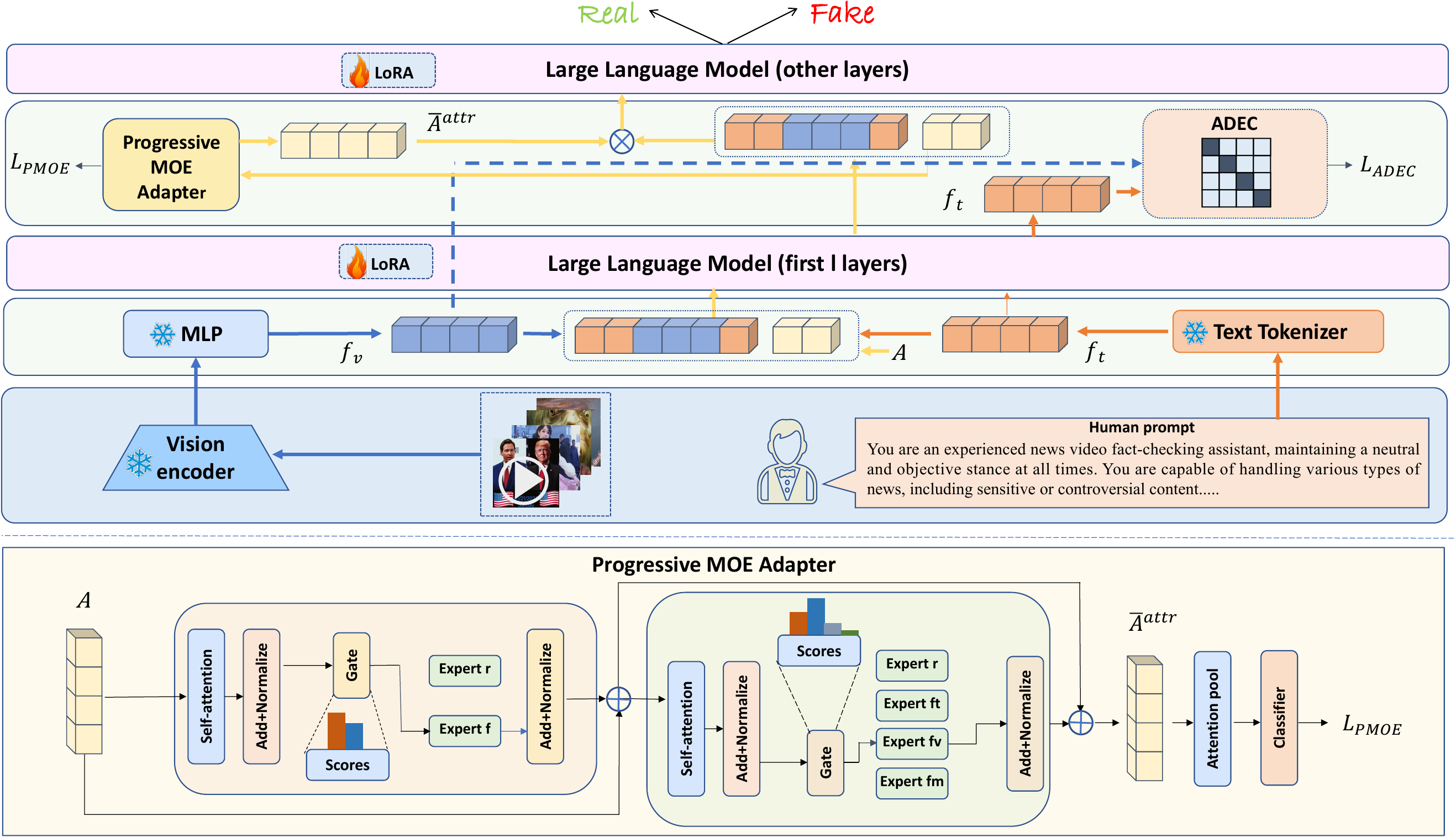} 
  \caption{Illustration of the proposed FakeSV-VLM framework, which is built upon a Vision Language Model (VLM). The model is trained with the classification loss, together with objectives from PMOE and ADEC modules.
}
  \label{Fig2}
\end{figure*}

\subsection{Overview}
Our FakeSV-VLM architecture is shown in Figure~\ref{Fig2}. During training, the model takes as input keyframes and textual descriptions, which are encoded by the visual encoder and tokenizer of the VLM. These multimodal features are concatenated with learnable Artifact Tokens and passed into the early layers of the LLM. The updated Artifact Tokens are then fed into the Progressive MoE Adapter (PMOE) for two-stage reasoning: overall authenticity prediction and manipulation type attribution. Meanwhile, the Alignment-driven Event Checking (ADEC) module aligns semantic information across modalities via contrastive learning. Ultimately, the entire model is optimized using the final classification loss, together with the losses from the PMOE and ADEC modules.

\subsection{Event-Prompting Template Design}
\paragraph{Prompt.}
To incorporate role-specific and contextual information into the Vision Language Model (VLM), we begin by designing a prompt grounded in prior work. Specifically, we refer to the prompting format used in the comparative experiments of \cite{r9}, which serves as a foundation for our design. Additional details are provided in Appendix~\ref{B}. The initial system prompt is as follows:

\begin{quote}
\textit{You are an experienced news video fact-checking assistant, maintaining a neutral and objective stance at all times. You are capable of handling various types of news, including sensitive or controversial content.}
\end{quote}

Building upon this, and given the description and keyframes of the input text, we further refine the VLM-specific prompt by explicitly incorporating metadata such as the news event, which has often been overlooked in previous work \cite{r9}. This metadata serves as a valuable pre-example to enhance the understanding and judgment of the model. The revised prompt is shown below:

\begin{quote}
\textit{Given the news video description ,new event and key frames, you need to predict the authenticity of the news video. If the video is more likely to be fake news, return fake; otherwise, return real. Please avoid providing ambiguous evaluations such as undetermined.}\\
\textit{News video description: <description>} \\ 
\textit{News events: <event>} \\ 
\textit{News video key frames: <video>} \\ 
\textit{Your prediction (no need to give your analysis, return real or fake only):}
\end{quote}

Finally, we concatenate the two prompts to form the complete human prompt fed into the VLM.

\paragraph{Answer.}
Subsequently, the VLM generates a binary response based on the constructed prompt, determining the authenticity of the input news videos as either \textit{real} or \textit{fake}.

\subsection{Progressive MoE Adapter}

Given a news video represented by a sequence of keyframes \(v = \{v_1, v_2, \dots, v_k\}\), a human prompt \(t\), a connector \(\mathcal{M}\), a tokenizer \(\mathcal{T}_k\) and a large language model (LLM), we first extract the visual features using a vision encoder \(\mathcal{E}_v\). These features are then mapped to the input space of the LLM via the connector, resulting in \(f_v = \mathcal{M}(\mathcal{E}_v(v))\). In parallel, we obtain the textual embedding of the prompt using the tokenizer, defined as \(f_t = \mathcal{T}_k(t)\).

\paragraph{Artifact Tokens.} 

To facilitate forgery-aware representation learning, we introduce a set of \(q\) learnable Artifact Tokens, denoted as \(A\), which are designed to capture potential manipulation signals in multimodal content. These tokens are jointly optimized during training to attend to cross-modal inconsistencies, such as semantic mismatches between visual and textual information or visual-level forgeries.

Specifically, we construct a fused input by concatenating the visual features \(f_v\) and the textual features \(f_t\) via \(\otimes\), where \(\otimes\) denotes the concatenation operation. The fused multimodal features \(f_c = f_v \otimes f_t\) are then augmented with the Artifact Tokens \(A\), and the combined representation \(f_c \otimes A\) is passed into the first \(l\) layers of the LLM to derive contextualized embeddings:
\begin{equation}
f_c \otimes A = \mathrm{LLM}_{1:l}(f_c \otimes A).
\label{eq1}
\end{equation}
For brevity, we still use $f_c$ and $A$ to represent the resultant features.

\paragraph{Detection MoE.}  

To perform an initial assessment of the authenticity of a news video, we introduce the Detection MoE module as the first decision-making stage. This module is designed to determine whether the input video is real or fake based on the contextualized Artifact Tokens \(A\).

The design of the Detection MoE follows the sparse activation paradigm proposed in~\cite{r35}. It consists of a multi-query attention layer, two layer normalization operations, a learnable gating mechanism, and two expert MLPs. Each expert network is implemented as a two-layer linear projection with a GELU activation function to capture nonlinear patterns from the token representations. The gating mechanism applies a linear transformation followed by a softmax operation to compute a token-wise distribution over the two experts, enabling dynamic routing based on token semantics.

For the contextualized Artifact Tokens \( A \) obtained from the previous stage, each token is dynamically routed to the most appropriate expert network based on its semantic information. Specifically, we first apply multi-head attention to the tokens as follows:
\begin{equation}
\bar{A}= \mathrm{softmax} \left( \frac{(A W_q)(A W_k)^\top}{\sqrt{d_k}} \right)\cdot(A W_v), \\
\label{eq2}
\end{equation}
where \( W_q \) denotes query projection matrix specific to each attention head, while \( W_k \) and \( W_v \) are the shared key and value projection matrices across all heads. \( d_k \) is the dimensionality of the key vectors.

Next, for the \(i\)-th token \( \bar{A}_i \), we assume the routing probabilities are \( p^r_i \) and \( p^f_i \), corresponding to the likelihood of being assigned to the corresponding expert, respectively. The expert with the highest routing probability is then selected as:
\begin{equation}
z = \arg\max_{\{r, f\}} \{ p^r_i, p^f_i \}.  
\label{eq3}
\end{equation}

Once the expert is selected, the attention-refined token representation is further processed through the corresponding expert network:
\begin{equation}
\begin{aligned}
\bar{A}_i^{\text{det}} = \mathrm{LN} \Big( \mathrm{Exp}_z\big( \mathrm{LN} \mathrm(\bar{A}_i\big) \Big)+\bar{A}_i,
\label{eq4}
\end{aligned}
\end{equation}
where \( \mathrm{LN}(\cdot) \) denotes the layer normalization operation, and \( \mathrm{Exp}_z(\cdot) \) represents the selected expert network, implemented as two linear layers with a GELU activation function.

Since the annotation of label or fake is available during training, we further augment the Detection MoE with an \emph{Authenticity Probability
Guidance (APG)}. 
In specific, the final authenticity probability \( p_r \) and \( p_f \), which represent the overall likelihood that the input news video is real or fake, respectively. This module is trained with a binary cross-entropy loss, defined as:
\begin{align}
p_r &= \frac{1}{q}\sum_{i}p_i^r, \quad 
p_f = \frac{1}{q}\sum_{i}p_i^f, \label{eq5} \\
L_{\textit{APG}} &= - \left( (1 - y) \log p_r + y \log p_f \right) \label{eq6},
\end{align}
where \( y \) denotes the ground-truth label.

\paragraph{Attribution MoE.}

To further identify fine-grained manipulation types in news videos, we feed the contextualized artifact representation \( \bar{A}^{\text{det}} \), obtained from the Detection MoE module, into the Attribution MoE module for more fine-grained forgery classification.

Unlike Detection MoE, Attribution MoE targets four manipulation scenarios: both video and text are real, only text is fake, only video is fake, and both are fake. It shares the same architecture as Detection MoE, consisting of a multi-query self-attention layer, two layer normalization layers, a learnable gating module, and four expert MLPs. For consistency, we still denote the Artifact Tokens after attention as \( \bar{A}^{\text{det}} \).

Similarly, for the \(i\)-th token \( \bar{A}^{\text{det}}_i \), we denote its routing probabilities to the four experts as \(p^r_i\), \(p^{fv}_i\), \(p^{ft}_i\), and \(p^{fm}_i\), which respectively represent the likelihood that this token belongs to each expert at a finer granularity level. The expert with the highest routing probability is then selected as:
\begin{equation}
z = \mathop{\mathrm{argmax}}_{\{r, fv, ft, fm\}} \left\{ p^r_i,\ p^{fv}_i, p^{ft}_i, p^{fm}_i \right\}.
\label{eq7}
\end{equation}

Once the expert is selected, the corresponding expert network is applied to the token \(A_i\), yielding the final output of this module:
\begin{equation}
\bar{A}_i^{\text{attr}} = \mathrm{LN} \left( \mathrm{Exp}_z \left( \mathrm{LN}(\bar{A}_i^{\text{det}}) \right) \right) + \bar{A}_i^{\text{det}}.
\label{eq8}
\end{equation}

\paragraph{Manipulation-Guided Artifact Perceiving.}  
To further enhance the discriminative capability of the Artifact Tokens, we introduce the Manipulation-Guided Artifact Perceiving process. The input to this module is the fused representation \( \bar{A}^{\text{attr}} \) from the Attribution MoE, which is used to guide attention toward potential manipulation patterns.

Specifically, we first apply attention pooling: a linear layer is used to compute attention scores for each token, which are then normalized by a softmax function to obtain token-wise weights. These weighted features are aggregated and passed through a two-layer MLP, the final output consists of two confidence scores, indicating the probabilities that the input video is real or fake, respectively, denoted as \( p'_{\text{r}} \) and \( p'_{\text{f}} \), respectively. The computation is defined as:
\begin{align}
p'_{\text{r}}, p'_{\text{f}}&=\mathrm{MLP}(\sum_{i=1}^{q} w_i \cdot \bar{A}_i^{\text{attr}}), \label{eq9}\\
w&=\mathrm{softmax} \big( \mathrm{FC}(\bar{A}^{\text{attr}})),
\label{eq10}
\end{align}
where \(\mathrm{FC}(\cdot)\) denotes a linear projection, \(\mathrm{MLP}(\cdot)\) denotes a two-layer multilayer perceptron, and \( w \in \mathbb{R}^{1 \times q} \) denote the weight vector, where \( q \) is the number of Artifact Tokens.

We supervise this module using the \emph{Artifact Classification Loss (ACL)} against the ground-truth label \( y \):
\begin{equation}
\mathcal{L}_{\textit{ACL}} = - \left( (1 - y) \log p'_{\text{r}} + y \log p'_{\text{f}} \right).
\label{eq11}
\end{equation}
This mechanism guides Artifact Tokens to attend to manipulation cues, thus improving the model’s capability to detect forged content.

To optimize the two-stage reasoning in PMOE, the total training loss is defined as:
\begin{equation}
\mathcal{L}_{\textit{PMOE}} = \frac{\mathcal{L}_{\textit{APG}} + \mathcal{L}_{\textit{ACL}}}{2}.
\label{eq12}
\end{equation}

This progressive optimization enables the model to first assess the overall authenticity of the news video, and then further infer the specific type of manipulation, enhancing robustness and generalization under multimodal conditions.

\paragraph{Answer Decoding.}  

Finally, the output from the PMOE module \( \bar{A}^{\text{attr}} \) is concatenated with the multimodal contextual representation \( f_c \), and the combined features are passed to the remaining layers of the LLM for decoding:
\begin{equation}
P = \mathrm{LLM}_{l+1:L} \left( f_c \otimes \bar{A}^{\text{attr}} \right).
\label{eq13}
\end{equation}
The output logits \( P \) are used to compute the final classification loss \( \mathcal{L}_{\textit{CE}} \) with respect to the ground-truth label.

\subsection{Alignment-driven Event Checking}

To preceive the inconsistency between visual and textual modalities at the event level in fake news videos, we introduce the Alignment-driven Event Checking (ADEC) module to help the model learn modality-consistent representations. Specifically, given the previously obtained text features \(f_t\) and visual features \(f_v\), we first feed the textual features \(f_t\) into the first \(l\) layers of a Large Language Model (LLM) to obtain contextualized text embeddings:
\begin{equation}
f_t = \mathrm{LLM}_{1:l}(f_t).
\label{eq14}
\end{equation}
For brevity, we still use $f_t$ to represent the resultant features.

We then perform average pooling on the visual features \( f_v \) and the contextualized text features \( f_t \) to extract representative global information \( \bar{f_v} \) and \( \bar{f_t} \). The pooled features \( \bar{f_v} \) and \( \bar{f_t} \) are then used in a contrastive learning setup to align semantic content across modalities. We compute the modality matching scores for both directions:
\begin{align}
s_{ij}^{v \to t} &= \frac{\exp(\mathrm{sim}(\bar{f_i^v}, \bar{f_j^t})/\tau)}{\sum_{j=1}^N \exp(\mathrm{sim}(\bar{f_i^v}, \bar{f_j^t})/\tau)}, \label{eq15}\\
s_{ij}^{t \to v} &= \frac{\exp(\mathrm{sim}(\bar{f_i^t}, \bar{f_j^v})/\tau)}{\sum_{j=1}^N \exp(\mathrm{sim}(\bar{f_i^t}, \bar{f_j^v})/\tau)}, \label{eq16}
\end{align}
where \( \mathrm{sim}(\cdot, \cdot) \) denotes cosine similarity, \( \tau \) is the temperature parameter, \( N \) is the batch size, and \( \bar{f_i^v} \) and \( \bar{f_i^t} \) denote the \( i \)-th news video and corresponding text in a batch. 

Subsequently, to guide the model in learning semantically aligned cross-modal representations, we construct a symmetric contrastive loss based on the previously computed modality matching scores to optimize the training process:
\begin{align}
\mathcal{L}_{v \to t} &= - \frac{1}{N} \sum_{i=1}^N \sum_{j=1}^N \mathcal{I}(v, t) \log s_{ij}^{v \to t}, \label{eq17}\\
\mathcal{L}_{t \to v} &= - \frac{1}{N} \sum_{i=1}^N \sum_{j=1}^N \mathcal{I}(v, t) \log s_{ij}^{t \to v}, \label{eq18}
\end{align}
where \(\mathcal{I}(v, t)\) denotes the match label, set to matched if the video and text belong to the same news and the news is real, and unmatched otherwise, including cross-news pairs and fake samples.

The final \emph{Alignment-driven Event Checking (ADEC)} loss is given by:
\begin{equation}
\mathcal{L}_{\textit{ADEC}} = \frac{\mathcal{L}_{v \to t} + \mathcal{L}_{t \to v}}{2}.
\label{eq19}
\end{equation}

\subsection{Training and Inference}

\paragraph{Training.}

The overall training objective combines the losses from the Progressive MoE Adapter (\( \mathcal{L}_{\textit{PMOE}} \)), the Alignment-driven Event Checking (\( \mathcal{L}_{\textit{ADEC}} \)), and the final classification stage (\( \mathcal{L}_{\textit{CE}} \)). The total loss is defined as:
\begin{equation}
\mathcal{L}_{\textit{total}} = \mathcal{L}_{\textit{CE}} + \mathcal{L}_{\textit{PMOE}} + \mathcal{L}_{\textit{ADEC}}.
\label{eq20}
\end{equation} 

\paragraph{Inference.}

It is worth noting that during inference, the expert-based decision process in the PMOE module and the contrastive alignment in the ADEC module are no longer required. The model can directly utilize the fine-tuned Vision Language Model (VLM) to generate authenticity predictions for news videos, thereby avoiding any additional computational overhead.

\section{Experiments}
\label{Experiments}

\subsection{Experiments setup}

\subsubsection{Implementation Details}

In this experiment, all results are obtained using 4 NVIDIA GeForce RTX 4090 GPUs. To maintain consistency, we evenly sample 8 frames from each video based on the video’s duration, with each frame resized to $448 \times 448$. We use InternVL2.5-8B \cite{r20} as the backbone VLM and select LoRA \cite{r8} for fine-tuning, with the LoRA rank set to 8 and LoRA alpha set to 32, using bfloat16 precision. The temperature \( \tau \) used in Eq.\ref{eq15}, \ref{eq16} is set to 0.07. Additionally, our batch size is set to 4, we train for 5 epochs on each dataset and select the parameters that perform best on the validation set for testing. During the first tenth of the total training steps, the learning rate is warmed up to 8e-5, and then a cosine schedule is used to decay the learning rate. We use the AdamW optimizer to update the parameters, with a weight decay of 0.1.

\subsubsection{Datasets and Evaluation Metrics}

\paragraph{Dataset.}
To evaluate the effectiveness of our FakeSV-VLM, we conduct experiments on two real-world fake news video datasets: FakeSV \cite{r12} and FakeTT \cite{r9}. The dataset partitioning follows prior work  \cite{r9}, \cite{r11}, \cite{r10}, \cite{r12}, dividing the dataset based on the release time of news videos with a 70\%, 15\%, and 15\% split for the training, validation, and test sets, respectively. For more details about the datasets, please refer to the Appendix \ref{C.1}.

\paragraph{Evaluation Metrics.}
We follow previous work \cite{r9}, \cite{r11}, \cite{r10}, \cite{r12} and use four metrics to evaluate the model's performance: Accuracy (ACC), Macro F1 score (M-F1), Macro Precision (M-P), and Macro Recall (M-R).

\subsection{Quantitative Results}

\begin{table*}[t]
  \centering
  \begin{tabular}{lcccc|cccc}
    \toprule[1.5pt]
    \textbf{Dataset} & \multicolumn{4}{c}{\textbf{FakeSV}} & \multicolumn{4}{c}{\textbf{FakeTT}} \\
    \hline
    \textbf{Model} & ACC & M-F1 & M-P & M-R & ACC & M-F1 & M-P & M-R \\
    \hline
    GPT-4o-mini~\nocite{r17} & 68.08 & 68.05 & 69.88 & 69.49 & 61.54 & 61.20 & 64.41 & 65.89 \\
    GPT-4.1-mini~\nocite{r18} & 70.30 & 70.25 & 70.61 & 70.87 & 49.16 & 48.54 & 62.50 & 59.70 \\
    Qwen2.5-VL~\nocite{r19} & 64.21 & 60.79 & 64.55 & 61.52 & 45.82 & 45.31 & 56.69 & 55.42 \\
    InternVL2.5~\nocite{r20} & 64.39 & 57.89 & 68.52 & 60.50 & 46.82 & 45.29 & 64.92 & 59.23 \\
    InternVL2.5-MPO~\nocite{r20} & 65.13 & 61.07 & 66.46 & 62.12 & 43.14 & 40.84 & 61.90 & 56.23 \\
    \hline
    ViT~\nocite{r13} & 70.85 & 70.66 & 70.64 & 70.91 & 64.88 & 62.59 & 62.54 & 63.80 \\
    Bert~\nocite{r14} & 78.41 & 78.25 & 78.17 & 78.52 & 70.90 & 69.00 & 68.71 & 70.60 \\
    \hline
    TikTec~\nocite{r15} & 73.06 & 72.79 & 72.73 & 72.93 & 66.56 & 65.55 & 66.50 & 68.62 \\
    FANVM~\nocite{r16} & 79.88 & 78.91 & 80.98 & 78.42 & 71.91 & 70.85 & 71.21 & 73.90 \\
    SV-FEND~\nocite{r12} & 80.81 & 80.19 & 81.08 & 79.84 & 77.26 & 75.55 & 74.94 & 77.13 \\
    FakingRecipe~\nocite{r19} & 84.69 & 84.39 & 84.57 & 84.25 & 79.26 & 77.53 & 76.86 & 78.89 \\
    CA-FVD~\nocite{r11} & 85.79 & 85.28 & 86.57 & 84.78 & 81.61 & 80.26 & 79.50 & 82.17 \\
    ExMRD~\nocite{r10} & 86.90 & 86.52 & 87.31 & 86.13 & 84.28 & 83.13 & 82.27 & 85.19 \\
    \hline
    \textbf{FakeSV-VLM (Ours)} & \textbf{90.22} & \textbf{89.97} & \textbf{90.55} & \textbf{89.64} & \textbf{89.30} & \textbf{87.98} & \textbf{87.80} & \textbf{88.17} \\
    \bottomrule[1.5pt]
  \end{tabular}
  \caption{Performance comparison on two datasets. Best results are shown in \textbf{bold}.}
  \label{table1}
\end{table*}

\begin{table*}[t]
  \centering
  \begin{tabular}{ccccc|cccc|cccc}
    \toprule[1.5pt]
    \multicolumn{5}{c|}{\textbf{Dataset}} & \multicolumn{4}{c|}{\textbf{FakeSV}} & \multicolumn{4}{c}{\textbf{FakeTT}} \\
    \hline
    \textbf{A} & \textbf{B} & \textbf{C} & \textbf{D} & \textbf{E} & ACC & M-F1 & M-P & M-R & ACC & M-F1 & M-P & M-R \\
    \hline
     &  &  &  & & 88.38 & 88.11 & 88.51 & 87.86 & 86.62 & 85.26 & 84.58 & 86.17 \\
    \checkmark &  &  & \checkmark & & 89.67 & 89.45 & 89.75 & 89.24 & 87.96 & 86.67 & 86.07 & 87.43 \\
     & \checkmark &  & \checkmark & & 89.85 & 89.64 & 89.92 & 89.45 & 88.29 & 87.01 & 86.47 & 87.68 \\
     &  & \checkmark & \checkmark & & 89.48 & 89.26 & 89.59 & 89.03 & 88.29 & 86.75 & 86.85 & 86.66 \\
     & \checkmark & \checkmark & \checkmark & & 90.04 & 89.79 & 90.32 & 89.48 & 88.63 & 87.35 & 86.87 & 87.93 \\
     & \checkmark & \checkmark &  & \checkmark & 88.75 & 88.29 & 90.07 & 87.64 & 87.96  & 87.01 & 86.01 & \textbf{88.96}  \\
    \hline
     & \checkmark & \checkmark & \checkmark & \checkmark & \textbf{90.22} & \textbf{89.97} & \textbf{90.55} & \textbf{89.64} & \textbf{89.30} & \textbf{87.98} & \textbf{87.80} & 88.17 \\
    \bottomrule[1.5pt]
  \end{tabular}
  \caption{Ablation study of components in FakeSV-VLM. \textbf{A}: Detection MoE without \emph{Authenticity Probability Guidance}; \textbf{B}: Detetction MoE; \textbf{C}: Attribution  MoE; \textbf{D}: Manipulation-Guided Artifact Perceiving; \textbf{E}: ADEC.}
  \label{table2}
\end{table*}

To comprehensively evaluate the performance of our method, we conduct comparisons with 13 competitive baselines. For more details about the baselines, please refer to the Appendix~\ref{C.2}. Extensive experiments are performed on two real-world datasets: FakeSV and FakeTT. To ensure the reliability of the results, each experiment is repeated three times, and the average performance is reported. The detailed quantitative results are presented in Table~\ref{table1}. Based on the results, we can make the following observations:

First, FakeSV-VLM outperforms all existing comparison methods across all evaluation metrics. Specifically, it achieves improvements of 3.32\% and 5.02\% in Accuracy, and 3.45\% and 4.85\% in Macro-F1 on the FakeSV and FakeTT datasets respectively, demonstrating the effectiveness of the proposed model.

Second, Compared to unimodal methods, multimodal approaches achieve average improvements of 8.42\% and 10.71\% in Accuracy on the FakeSV and FakeTT datasets, respectively, highlighting the importance of jointly modeling visual and textual features in fake news videos detection. FakeSV-VLM effectively leverages multimodal information and achieves significant performance gains.

Finally, methods based on Vision-Language Models (VLMs) show notably lower performance than both unimodal and multimodal approaches, the average Accuracy on FakeSV and FakeTT is only 66.42\% and 49.30\%, indicating that zero-shot VLMs alone are insufficient for fake news video detection. Although closed-source VLMs like the GPT series generally outperform open-source ones, their advantages are not always consistent—e.g., GPT-4.1-mini performs poorly on the FakeTT dataset and tends to over-predict fakeness. In contrast, FakeSV-VLM builds on a lightweight open-source VLM with task-specific fine-tuning, leading to significantly improved performance over standard VLM-based zero-shot methods.

\subsection{Ablation Study}

\paragraph{Components Ablation Analysis.}

Table~\ref{table2} presents the contribution of each component in our framework. Compared to fine-tuning the backbone alone, incorporating the PMOE module yields notable performance gains on both datasets. Our model achieves over 90\% accuracy on FakeSV and surpasses 88\% on FakeTT for the first time. Furthermore, using either the Detection MoE or the Attribution MoE individually still brings considerable improvements, demonstrating the effectiveness of each detection branch. We also observe that even when used in isolation, the Manipulation-Guided Artifact Perceiving (MGAP) process helps enhance the model’s ability to identify manipulation cues, contributing to overall performance. Notably, when the Detection MoE does not leverage the \emph{Authenticity Probability Guidance}, a slight performance drop is observed, which further validates the necessity of this guidance mechanism. In addition, the ADEC module alone brings partial improvements on both datasets, highlighting the importance of event-level semantic alignment across modalities in fake news video detection.

\begin{figure*}[ht]
  \centering
  \includegraphics[width=\textwidth]{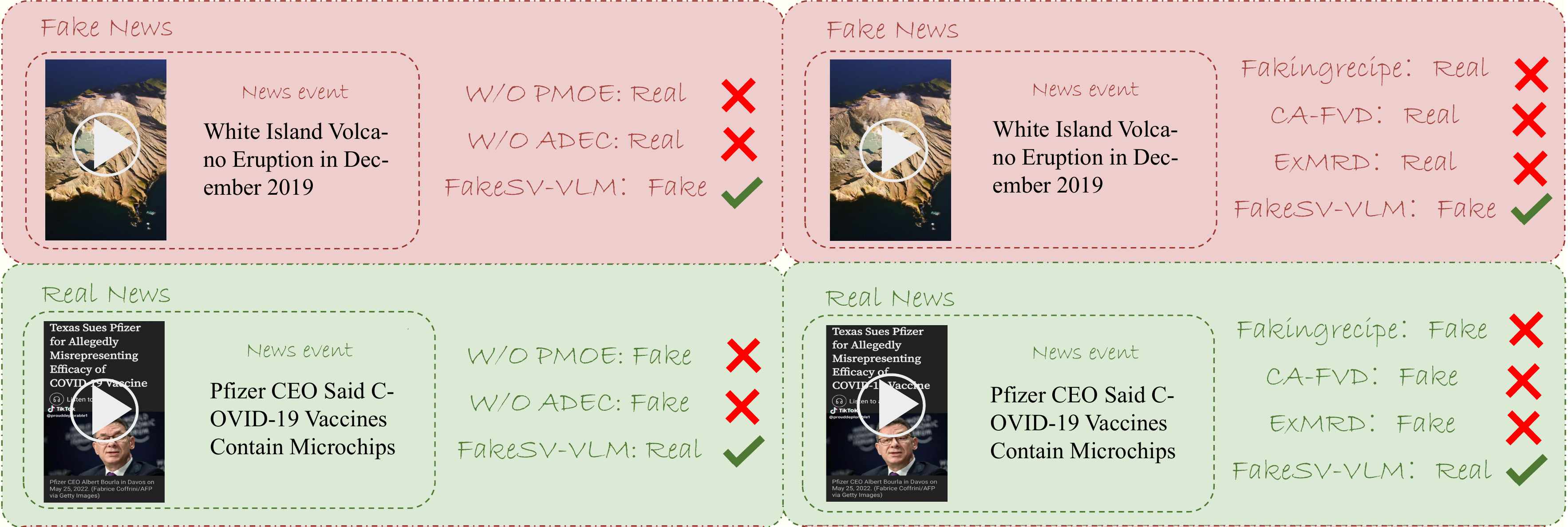} 
  \caption{Two classic cases of fake news videos selected from the FakeTT dataset.}
  \label{Fig4}
\end{figure*}

\begin{figure}[ht]
  \centering
  \begin{subfigure}[b]{0.48\textwidth}
    \includegraphics[width=\textwidth]{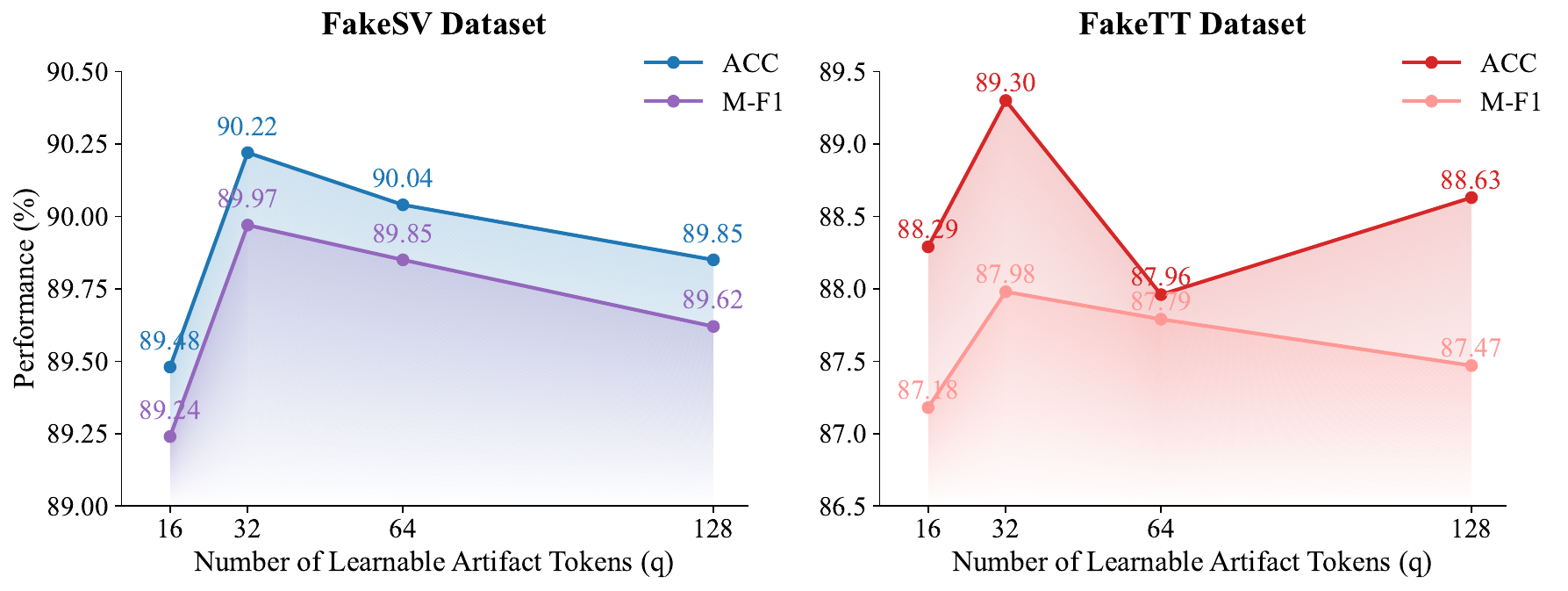}
    \caption{Impact of the number of Artifact Tokens \(q\).}
    \label{Fig3a}
  \end{subfigure}
  \hfill
  \begin{subfigure}[b]{0.48\textwidth}
    \includegraphics[width=\textwidth]{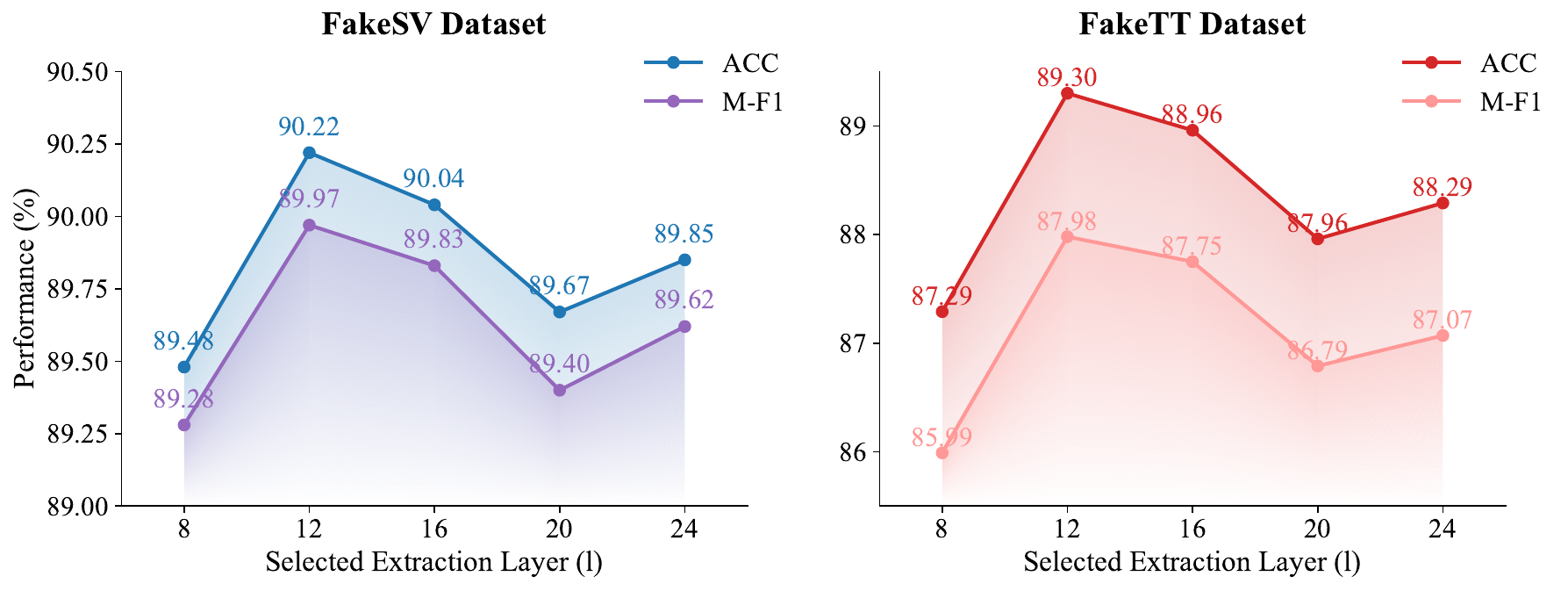}
    \caption{Impact of the extraction layer \(l\).}
    \label{Fig3b}
  \end{subfigure}
  \caption{Ablation studies of different architectural components in our FakeSV-VLM.}
  \label{Fig3}
\end{figure}

\paragraph{Impact of the Number of Artifact Tokens \textit{\textbf{q}}.}

We analyze the impact of the number of Artifact Tokens \( q \) on the final detection performance, as shown in Figure~\ref{Fig3a}. Experimental results demonstrate that the model achieves the best performance when \( q=32 \). When the number of tokens is too small, the model fails to capture sufficient key information across modalities. Conversely, an excessive number of tokens may introduce redundant information, which interferes with the model's representation learning and leads to performance degradation.

\paragraph{Selection of Layer \textit{\textbf{l}}.} 

We analyze the effect of extracting learnable Artifact Tokens and performing contrastive learning at different layers \( l \) of the LLM, as shown in Figure~\ref{Fig3b}. The total number of layers \(L\) is 32. Experimental results show that extracting tokens from shallower layers (e.g., layer 8) yields suboptimal performance, indicating that the model has not yet formed sufficiently rich semantic representations at these depths. In contrast, extracting tokens from intermediate layers (e.g., layers 12 and 16) significantly improves performance, as the model is better able to integrate and align multimodal information. However, performance begins to decline when extracting from deeper layers (e.g., layers 20 and 24), likely due to excessive abstraction, which causes the token representations to lose alignment-relevant information and ultimately degrades detection performance.

For more experimental results, please refer to the Appendix~\ref{C.3}, ~\ref{C.4} and ~\ref{C.5}.

\subsection{Case Studies}

As illustrated in Figure~\ref{Fig4}, we present two representative cases from the FakeTT dataset to demonstrate the effectiveness of our proposed FakeSV-VLM framework. 

The first case, titled \textit{"White Island Volcano Eruption in December 2019"}. Although the event is real, the video was repurposed from another eruption to support a misleading narrative. Existing models fail to detect this manipulation. PMOE infers authenticity and manipulation types, and ADEC reveals semantic inconsistencies across modalities. Together, they enable accurate classification of the video as fake news.

The t case concerns the event \textit{"Pfizer CEO Said COVID-19 Vaccines Contain Microchips"}. The video itself is authentic, but its content was misinterpreted to propagate a false narrative. When applied independently, PMOE mistakenly flags the video as manipulated, while ADEC erroneously infers cross-modal inconsistencies. However, when combined, they complement each other and yield the correct classification of the video as genuine news.

Additional other cases can be found in the Appendix~\ref{D}.

\section{Conclusion}

This paper presents FakeSV-VLM, a novel framework for multimodal fake news videos detection that fully leverages the reasoning capabilities of Vision Language Models (VLMs) within an end-to-end training paradigm. FakeSV-VLM integrates VLMs into the entire optimization process, enabling more effective modeling of semantic authenticity. To this end, we design the PMOE module, which captures hierarchical manipulation patterns through a two-stage expert mechanism, and the ADEC module, which facilitates cross-modal semantic alignment via contrastive learning. Extensive experiments on real-world datasets demonstrate that FakeSV-VLM consistently outperforms state-of-the-art baselines, showcasing its strong generalizability and effectiveness.

\section*{Limitations}

Our framework leverages vision-language models (VLMs) to detect fake news videos by integrating multiple specialized components. Despite the strong performance, several limitations remain. First, due to limited computational resources, we are currently unable to scale our system to larger VLM backbones, which may further enhance performance. Second, the lack of high-quality, fine-grained benchmark datasets remains a challenge. In particular, there is no existing dataset that provides detailed annotations indicating which specific parts of a news video are fabricated or misleading. In future work, we plan to construct such a dataset to support more interpretable and fine-grained evaluation of multimodal misinformation detection systems.

\section*{Acknowledgment}

The computation is completed on the HPC Platform of Hefei University of Technology. This work is supported by the Fundamental Research Funds for the Central Universities (No. JZ2024HGTB0261) and the National Natural Science Foundation of China (NSFC) under Grant No. 62302140 and No. 62572166.

\bibliography{custom}

\begin{thebibliography}{44}
\providecommand{\natexlab}[1]{#1}

\bibitem[{Abbasi~Aghamaleki and Behrad(2017)}]{r3}
Javad Abbasi~Aghamaleki and Alireza Behrad. 2017.
\newblock Malicious inter-frame video tampering detection in mpeg videos using time and spatial domain analysis of quantization effects.
\newblock \emph{Multimedia Tools and Applications}, 76:20691--20717.

\bibitem[{Achiam et~al.(2023)Achiam, Adler, Agarwal, Ahmad, Akkaya, Aleman, Almeida, Altenschmidt, Altman, Anadkat et~al.}]{r17}
Josh Achiam, Steven Adler, Sandhini Agarwal, Lama Ahmad, Ilge Akkaya, Florencia~Leoni Aleman, Diogo Almeida, Janko Altenschmidt, Sam Altman, Shyamal Anadkat, and 1 others. 2023.
\newblock Gpt-4 technical report.
\newblock \emph{arXiv preprint arXiv:2303.08774}.

\bibitem[{Al-Sanjary et~al.(2018)Al-Sanjary, Ahmed, Jaharadak, Ali, and Zangana}]{r25}
Omar~Ismael Al-Sanjary, Ahmed~Abdullah Ahmed, Adam Amril~Bin Jaharadak, Musab~AM Ali, and Hewa~Majeed Zangana. 2018.
\newblock Detection clone an object movement using an optical flow approach.
\newblock In \emph{2018 IEEE Symposium on Computer Applications \& Industrial Electronics (ISCAIE)}, pages 388--394. IEEE.

\bibitem[{Bai et~al.(2025)Bai, Chen, Liu, Wang, Ge, Song, Dang, Wang, Wang, Tang et~al.}]{r19}
Shuai Bai, Keqin Chen, Xuejing Liu, Jialin Wang, Wenbin Ge, Sibo Song, Kai Dang, Peng Wang, Shijie Wang, Jun Tang, and 1 others. 2025.
\newblock Qwen2. 5-vl technical report.
\newblock \emph{arXiv preprint arXiv:2502.13923}.

\bibitem[{Bu et~al.(2024)Bu, Sheng, Cao, Qi, Wang, and Li}]{r9}
Yuyan Bu, Qiang Sheng, Juan Cao, Peng Qi, Danding Wang, and Jintao Li. 2024.
\newblock Fakingrecipe: Detecting fake news on short video platforms from the perspective of creative process.
\newblock In \emph{Proceedings of the 32nd ACM International Conference on Multimedia}, pages 1351--1360.

\bibitem[{Bunti{\'c} et~al.(2020)Bunti{\'c}, Dami{\'c}, and Gregani{\'c}}]{r6}
Luka Bunti{\'c}, Mate Dami{\'c}, and Dalibor Gregani{\'c}. 2020.
\newblock Impact of fake news on the global economy.
\newblock \emph{Trade perspectives}, pages 73--81.

\bibitem[{Chen et~al.(2024)Chen, Wang, Cao, Liu, Gao, Cui, Zhu, Ye, Tian, Liu et~al.}]{r20}
Zhe Chen, Weiyun Wang, Yue Cao, Yangzhou Liu, Zhangwei Gao, Erfei Cui, Jinguo Zhu, Shenglong Ye, Hao Tian, Zhaoyang Liu, and 1 others. 2024.
\newblock Expanding performance boundaries of open-source multimodal models with model, data, and test-time scaling.
\newblock \emph{arXiv preprint arXiv:2412.05271}.

\bibitem[{Choi and Ko(2021)}]{r16}
Hyewon Choi and Youngjoong Ko. 2021.
\newblock Using topic modeling and adversarial neural networks for fake news video detection.
\newblock In \emph{Proceedings of the 30th ACM international conference on information \& knowledge management}, pages 2950--2954.

\bibitem[{Devlin et~al.(2019)Devlin, Chang, Lee, and Toutanova}]{r14}
Jacob Devlin, Ming-Wei Chang, Kenton Lee, and Kristina Toutanova. 2019.
\newblock Bert: Pre-training of deep bidirectional transformers for language understanding.
\newblock In \emph{Proceedings of the 2019 conference of the North American chapter of the association for computational linguistics: human language technologies, volume 1 (long and short papers)}, pages 4171--4186.

\bibitem[{Diao et~al.(2025{\natexlab{a}})Diao, Cheng, Barrios, and Jin}]{r38}
Xingjian Diao, Ming Cheng, Wayner Barrios, and SouYoung Jin. 2025{\natexlab{a}}.
\newblock Ft2tf: First-person statement text-to-talking face generation.
\newblock In \emph{Proceedings of the Winter Conference on Applications of Computer Vision (WACV)}, pages 4821--4830.

\bibitem[{Diao et~al.(2025{\natexlab{b}})Diao, Yang, Zhang, Wu, Cheng, and Gui}]{r40}
Xingjian Diao, Tianzhen Yang, Chunhui Zhang, Weiyi Wu, Ming Cheng, and Jiang Gui. 2025{\natexlab{b}}.
\newblock Learning sparsity for effective and efficient music performance question answering.
\newblock \emph{arXiv preprint arXiv:2506.01319}.

\bibitem[{Diao et~al.(2025{\natexlab{c}})Diao, Zhang, Kong, Wu, Ma, Ouyang, Qing, Vosoughi, and Gui}]{r42}
Xingjian Diao, Chunhui Zhang, Keyi Kong, Weiyi Wu, Chiyu Ma, Zhongyu Ouyang, Peijun Qing, Soroush Vosoughi, and Jiang Gui. 2025{\natexlab{c}}.
\newblock Soundmind: Rl-incentivized logic reasoning for audio-language models.
\newblock \emph{arXiv preprint arXiv:2506.12935}.

\bibitem[{Diao et~al.(2024)Diao, Zhang, Wu, Cheng, Ouyang, Wu, and Gui}]{r41}
Xingjian Diao, Chunhui Zhang, Tingxuan Wu, Ming Cheng, Zhongyu Ouyang, Weiyi Wu, and Jiang Gui. 2024.
\newblock "learning musical representations for music performance question answering".
\newblock In \emph{"Findings of the Association for Computational Linguistics: EMNLP 2024"}.

\bibitem[{Diao et~al.(2025{\natexlab{d}})Diao, Zhang, Wu, Ouyang, Qing, Cheng, Vosoughi, and Gui}]{r39}
Xingjian Diao, Chunhui Zhang, Weiyi Wu, Zhongyu Ouyang, Peijun Qing, Ming Cheng, Soroush Vosoughi, and Jiang Gui. 2025{\natexlab{d}}.
\newblock Temporal working memory: Query-guided segment refinement for enhanced multimodal understanding.
\newblock \emph{arXiv preprint arXiv:2502.06020}.

\bibitem[{Dosovitskiy et~al.(2020)Dosovitskiy, Beyer, Kolesnikov, Weissenborn, Zhai, Unterthiner, Dehghani, Minderer, Heigold, Gelly et~al.}]{r13}
Alexey Dosovitskiy, Lucas Beyer, Alexander Kolesnikov, Dirk Weissenborn, Xiaohua Zhai, Thomas Unterthiner, Mostafa Dehghani, Matthias Minderer, Georg Heigold, Sylvain Gelly, and 1 others. 2020.
\newblock An image is worth 16x16 words: Transformers for image recognition at scale.
\newblock \emph{arXiv preprint arXiv:2010.11929}.

\bibitem[{Fedus et~al.(2022)Fedus, Zoph, and Shazeer}]{r35}
William Fedus, Barret Zoph, and Noam Shazeer. 2022.
\newblock Switch transformers: Scaling to trillion parameter models with simple and efficient sparsity.
\newblock \emph{Journal of Machine Learning Research}, 23(120):1--39.

\bibitem[{Hong et~al.(2025)Hong, Lang, Xu, Cheng, Zhong, and Zhou}]{r10}
Rongpei Hong, Jian Lang, Jin Xu, Zhangtao Cheng, Ting Zhong, and Fan Zhou. 2025.
\newblock Following clues, approaching the truth: Explainable micro-video rumor detection via chain-of-thought reasoning.
\newblock In \emph{Proceedings of the ACM on Web Conference 2025}, pages 4684--4698.

\bibitem[{Hu et~al.(2024)Hu, Sheng, Cao, Shi, Li, Wang, and Qi}]{r34}
Beizhe Hu, Qiang Sheng, Juan Cao, Yuhui Shi, Yang Li, Danding Wang, and Peng Qi. 2024.
\newblock Bad actor, good advisor: Exploring the role of large language models in fake news detection.
\newblock In \emph{Proceedings of the AAAI Conference on Artificial Intelligence}, volume~38, pages 22105--22113.

\bibitem[{Hu et~al.(2022)Hu, Shen, Wallis, Allen-Zhu, Li, Wang, Wang, Chen et~al.}]{r8}
Edward~J Hu, Yelong Shen, Phillip Wallis, Zeyuan Allen-Zhu, Yuanzhi Li, Shean Wang, Lu~Wang, Weizhu Chen, and 1 others. 2022.
\newblock Lora: Low-rank adaptation of large language models.
\newblock \emph{ICLR}, 1(2):3.

\bibitem[{Li et~al.(2023)Li, Li, Savarese, and Hoi}]{r27}
Junnan Li, Dongxu Li, Silvio Savarese, and Steven Hoi. 2023.
\newblock Blip-2: Bootstrapping language-image pre-training with frozen image encoders and large language models.
\newblock In \emph{International conference on machine learning}, pages 19730--19742. PMLR.

\bibitem[{Li et~al.(2025{\natexlab{a}})Li, Cao, He, Cheng, Fu, Xiao, Wang, and Tang}]{r37}
Yanshu Li, Yi~Cao, Hongyang He, Qisen Cheng, Xiang Fu, Xi~Xiao, Tianyang Wang, and Ruixiang Tang. 2025{\natexlab{a}}.
\newblock \href {https://arxiv.org/abs/2504.04633} {M$^2$iv: Towards efficient and fine-grained multimodal in-context learning via representation engineering}.
\newblock \emph{Preprint}, arXiv:2504.04633.

\bibitem[{Li et~al.(2025{\natexlab{b}})Li, Yun, Yang, Feng, Huang, and Tang}]{r36}
Yanshu Li, Tian Yun, Jianjiang Yang, Pinyuan Feng, Jinfa Huang, and Ruixiang Tang. 2025{\natexlab{b}}.
\newblock Taco: Enhancing multimodal in-context learning via task mapping-guided sequence configuration.
\newblock \emph{arXiv preprint arXiv:2505.17098}.

\bibitem[{Lian et~al.(2024)Lian, Liu, Wang, Wu, Zhu, and Zheng}]{r44}
Jingchun Lian, Lingyu Liu, Yaxiong Wang, Yujiao Wu, Li~Zhu, and Zhedong Zheng. 2024.
\newblock \href {https://doi.org/10.48550/ARXIV.2412.19685} {A large-scale interpretable multi-modality benchmark for facial image forgery localization}.
\newblock \emph{CoRR}, abs/2412.19685.

\bibitem[{Liang et~al.(2024)Liang, Cai, Xu, Huang, Wang, Liang, Liu, Li, Wang, and Huang}]{r31}
Yaoyuan Liang, Zhuojun Cai, Jian Xu, Guanbo Huang, Yiran Wang, Xiao Liang, Jiahao Liu, Ziran Li, Jingang Wang, and Shao-Lun Huang. 2024.
\newblock Unleashing region understanding in intermediate layers for mllm-based referring expression generation.
\newblock \emph{Advances in Neural Information Processing Systems}, 37:120578--120601.

\bibitem[{Liu et~al.(2024)Liu, Li, Huang, Li, Cui, Liang, Qin, Deng, and He}]{r33}
Xuannan Liu, Peipei Li, Huaibo Huang, Zekun Li, Xing Cui, Jiahao Liang, Lixiong Qin, Weihong Deng, and Zhaofeng He. 2024.
\newblock Fka-owl: Advancing multimodal fake news detection through knowledge-augmented lvlms.
\newblock In \emph{Proceedings of the 32nd ACM International Conference on Multimedia}, pages 10154--10163.

\bibitem[{Niu et~al.(2023)Niu, Shrestha, Ghimire, and Lu}]{r7}
Shuo Niu, Dilasha Shrestha, Abhisan Ghimire, and Zhicong Lu. 2023.
\newblock A survey on watching social issue videos among youtube and tiktok users.
\newblock \emph{arXiv preprint arXiv:2310.19193}.

\bibitem[{OpenAI(2025)}]{r18}
OpenAI. 2025.
\newblock \href {https://openai.com/index/gpt-4-1/} {Introducing gpt-4.1 in the api}.
\newblock Accessed: 2025-05-01.

\bibitem[{Qi et~al.(2023{\natexlab{a}})Qi, Bu, Cao, Ji, Shui, Xiao, Wang, and Chua}]{r12}
Peng Qi, Yuyan Bu, Juan Cao, Wei Ji, Ruihao Shui, Junbin Xiao, Danding Wang, and Tat-Seng Chua. 2023{\natexlab{a}}.
\newblock Fakesv: A multimodal benchmark with rich social context for fake news detection on short video platforms.
\newblock In \emph{Proceedings of the AAAI Conference on Artificial Intelligence}, volume~37, pages 14444--14452.

\bibitem[{Qi et~al.(2024)Qi, Yan, Hsu, and Lee}]{r32}
Peng Qi, Zehong Yan, Wynne Hsu, and Mong~Li Lee. 2024.
\newblock Sniffer: Multimodal large language model for explainable out-of-context misinformation detection.
\newblock In \emph{Proceedings of the IEEE/CVF conference on computer vision and pattern recognition}, pages 13052--13062.

\bibitem[{Qi et~al.(2023{\natexlab{b}})Qi, Zhao, Shen, Ji, Cao, and Chua}]{r23}
Peng Qi, Yuyang Zhao, Yufeng Shen, Wei Ji, Juan Cao, and Tat-Seng Chua. 2023{\natexlab{b}}.
\newblock Two heads are better than one: Improving fake news video detection by correlating with neighbors.
\newblock \emph{arXiv preprint arXiv:2306.05241}.

\bibitem[{Saddique et~al.(2019)Saddique, Asghar, Bajwa, Hussain, and Habib}]{r24}
Mubbashar Saddique, Khurshid Asghar, Usama~Ijaz Bajwa, Muhammad Hussain, and Zulfiqar Habib. 2019.
\newblock Spatial video forgery detection and localization using texture analysis of consecutive frames.
\newblock \emph{Advances in Electrical \& Computer Engineering}, 19(3).

\bibitem[{Shang et~al.(2021)Shang, Kou, Zhang, and Wang}]{r15}
Lanyu Shang, Ziyi Kou, Yang Zhang, and Dong Wang. 2021.
\newblock A multimodal misinformation detector for covid-19 short videos on tiktok.
\newblock In \emph{2021 IEEE international conference on big data (big data)}, pages 899--908. IEEE.

\bibitem[{Sitara and Mehtre(2017)}]{r2}
K~Sitara and BM~Mehtre. 2017.
\newblock A comprehensive approach for exposing inter-frame video forgeries.
\newblock In \emph{2017 IEEE 13th International Colloquium on Signal Processing \& its Applications (CSPA)}, pages 73--78. IEEE.

\bibitem[{Venkatagiri et~al.(2023)Venkatagiri, Schafer, and Prochaska}]{r4}
Sukrit Venkatagiri, Joseph~S Schafer, and Stephen Prochaska. 2023.
\newblock The challenges of studying misinformation on video-sharing platforms during crises and mass-convergence events.
\newblock \emph{arXiv preprint arXiv:2303.14309}.

\bibitem[{Vykopal et~al.(2023)Vykopal, Pikuliak, Srba, Moro, Macko, and Bielikova}]{r21}
Ivan Vykopal, Mat{\'u}{\v{s}} Pikuliak, Ivan Srba, Robert Moro, Dominik Macko, and Maria Bielikova. 2023.
\newblock Disinformation capabilities of large language models.
\newblock \emph{arXiv preprint arXiv:2311.08838}.

\bibitem[{Wang et~al.(2025)Wang, Zhang, Wang et~al.}]{r11}
Junxi Wang, Na~Zhang, Yaxiong Wang, and 1 others. 2025.
\newblock Consistency-aware fake videos detection on short video platforms.
\newblock \emph{arXiv preprint arXiv:2504.21495}.

\bibitem[{Wang et~al.(2024)Wang, Bai, Tan, Wang, Fan, Bai, Chen, Liu, Wang, Ge et~al.}]{r28}
Peng Wang, Shuai Bai, Sinan Tan, Shijie Wang, Zhihao Fan, Jinze Bai, Keqin Chen, Xuejing Liu, Jialin Wang, Wenbin Ge, and 1 others. 2024.
\newblock Qwen2-vl: Enhancing vision-language model's perception of the world at any resolution.
\newblock \emph{arXiv preprint arXiv:2409.12191}.

\bibitem[{Wang et~al.(2019)Wang, Yang, Qian, Ma, Lu, Li, and Fan}]{r45}
Yaxiong Wang, Hao Yang, Xueming Qian, Lin Ma, Jing Lu, Biao Li, and Xin Fan. 2019.
\newblock \href {https://doi.org/10.24963/IJCAI.2019/526} {Position focused attention network for image-text matching}.
\newblock In \emph{Proceedings of the Twenty-Eighth International Joint Conference on Artificial Intelligence, {IJCAI} 2019, Macao, China, August 10-16,2019}, pages 3792--3798. ijcai.org.

\bibitem[{Wittenberg et~al.(2021)Wittenberg, Tappin, Berinsky, and Rand}]{r5}
Chloe Wittenberg, Ben~M Tappin, Adam~J Berinsky, and David~G Rand. 2021.
\newblock The (minimal) persuasive advantage of political video over text.
\newblock \emph{Proceedings of the National Academy of Sciences}, 118(47):e2114388118.

\bibitem[{Wu et~al.(2024)Wu, Guo, and Hooi}]{r22}
Jiaying Wu, Jiafeng Guo, and Bryan Hooi. 2024.
\newblock Fake news in sheep's clothing: Robust fake news detection against llm-empowered style attacks.
\newblock In \emph{Proceedings of the 30th ACM SIGKDD conference on knowledge discovery and data mining}, pages 3367--3378.

\bibitem[{Zampoglou et~al.(2019)Zampoglou, Markatopoulou, Mercier, Touska, Apostolidis, Papadopoulos, Cozien, Patras, Mezaris, and Kompatsiaris}]{r1}
Markos Zampoglou, Foteini Markatopoulou, Gregoire Mercier, Despoina Touska, Evlampios Apostolidis, Symeon Papadopoulos, Roger Cozien, Ioannis Patras, Vasileios Mezaris, and Ioannis Kompatsiaris. 2019.
\newblock Detecting tampered videos with multimedia forensics and deep learning.
\newblock In \emph{MultiMedia Modeling: 25th International Conference, MMM 2019, Thessaloniki, Greece, January 8--11, 2019, Proceedings, Part I 25}, pages 374--386. Springer.

\bibitem[{Zhang et~al.(2024{\natexlab{a}})Zhang, Wen, Wu, Qin, Xue', and Nie}]{r30}
Xian Zhang, Haokun Wen, Jianlong Wu, Pengda Qin, Hui Xue', and Liqiang Nie. 2024{\natexlab{a}}.
\newblock Differential-perceptive and retrieval-augmented mllm for change captioning.
\newblock In \emph{Proceedings of the 32nd ACM International Conference on Multimedia}, pages 4148--4157.

\bibitem[{Zhang et~al.(2025)Zhang, Wang, Wu, Wu, and Zhu}]{r43}
Yuchen Zhang, Yaxiong Wang, Yujiao Wu, Lianwei Wu, and Li~Zhu. 2025.
\newblock \href {https://doi.org/10.48550/ARXIV.2505.17476} {The coherence trap: When mllm-crafted narratives exploit manipulated visual contexts}.
\newblock \emph{CoRR}, abs/2505.17476.

\bibitem[{Zhang et~al.(2024{\natexlab{b}})Zhang, Wang, Cheng, Zhong, Guo, and Wang}]{r29}
Zhenxing Zhang, Yaxiong Wang, Lechao Cheng, Zhun Zhong, Dan Guo, and Meng Wang. 2024{\natexlab{b}}.
\newblock Asap: Advancing semantic alignment promotes multi-modal manipulation detecting and grounding.
\newblock \emph{arXiv preprint arXiv:2412.12718}.

\end{thebibliography}

\clearpage
\appendix

\section{Related Work}
\label{A}

\subsection{Fake News Videos Detection}

Fake news video detection aims to verify the authenticity of news events by analyzing video content and related metadata. Early methods focused on single modalities, such as analyzing texture relationships between frames \cite{r24} or changes in optical flow \cite{r25}, but often ignored other useful signals. Recent multimodal approaches have demonstrated greater effectiveness. For example, SV-FEND \cite{r12} integrates visual, textual, and audio information from video content, comments, and descriptions. ExMRD \cite{r10} further improves performance by distilling knowledge from VLMs to enrich text and aligning it with key video frames.

\subsection{Vision Language Models}

In recent years, multimodal learning has developed rapidly, demonstrating great potential in the integration and collaboration across different modalities such as vision, language, and speech \cite{r38,r39,r40,r41,r42,r44,r45}. As a prominent representative of this direction, Vision Language Models (VLMs) have attracted widespread attention for their ability to integrate multiple modalities. For instance, GPT-4 \cite{r17}, as a representative closed-source model, has made significant advances in both text and image understanding. Open-source VLMs have also thrived: BLIP2 \cite{r27} excels in image understanding and generation, while QwenVL2 \cite{r28} improves image and video comprehension via dynamic resolution. These models are widely applied to downstream tasks \cite{r29,r30,r31,r36,r37,r43}. In fake news detection, VLMs have also shown great promise \cite{r32,r33,r34}, though most focus on text-image modalities, with limited work on video. In this work, we incorporate VLMs into the training process to explore their potential in multimodal video-based fake news detection.

\section{Human Prompt Design}
\label{B}

To effectively guide the VLMs in performing judgment tasks, we assign it a well-defined role and design human-written prompts with contextual awareness. These prompts not only provide clear task instructions and maintain a consistent reasoning format, but also incorporate the linguistic styles of different data sources, enabling the model to better understand the task intent and generate responses that align with human expectations.

For the Chinese dataset FakeSV, we design Chinese prompts that align with the common linguistic habits and expressive styles of Chinese short-video platforms, while maintaining clarity and neutrality:

\begin{CJK*}{UTF8}{gbsn}

\begin{tcolorbox}[colback=gray!5!white, colframe=black!50, title=Prompt Used in FakeSV, sharp corners=southwest, fonttitle=\bfseries]
你是一名经验丰富的新闻视频事实核查助手，始终保持中立和客观的立场。你能够处理各种类型的新闻，包括敏感或有争议的内容。\\
根据新闻视频描述、新闻事件以及关键帧，你需要判断该新闻视频的真实性。如果该视频更可能是假新闻，请返回fake；否则，返回real。请避免使用诸如“无法确定”之类的模棱两可的评价。\\
新闻视频描述：<description> \\
新闻事件：<event> \\
新闻关键帧：<video> \\
你的判断（不需要分析，只返回real或fake）：
\end{tcolorbox}
\end{CJK*}

For the English dataset FakeTT, we construct English prompts using concise and direct language that mirrors the tone of user-generated short-video content, while preserving clarity and neutrality:

\begin{tcolorbox}[colback=gray!5!white, colframe=black!50, title=Prompt Used in FakeTT, sharp corners=southwest, fonttitle=\bfseries]
\textit{You are an experienced news video fact-checking assistant, maintaining a neutral and objective stance at all times. You are capable of handling various types of news, including sensitive or controversial content.} \\
\textit{Given the news video description, news event and key frames, you need to predict the authenticity of the news video. If the video is more likely to be fake news, return fake; otherwise, return real. Please avoid providing ambiguous evaluations such as undetermined.} \\
\textit{News video description: <description>} \\ 
\textit{News events: <event>} \\ 
\textit{News video key frames: <video>} \\ 
\textit{Your prediction (no need to give your analysis, return real or fake only):}
\end{tcolorbox}

\section{Other Experimental Details}
\label{C}

\subsection{Dataset}
\label{C.1}

To evaluate the effectiveness of our proposed method on multimodal fake news video detection, we utilize two real-world short video datasets: \textbf{FakeSV} for Chinese and \textbf{FakeTT} for English scenarios.

\paragraph{FakeSV.}
FakeSV is the largest Chinese short-video dataset for fake news detection at the time of release. It was collected from popular Chinese short-video platforms and covers widely circulated real and fake news content in authentic social environments. Each sample contains multiple modalities, including the video itself, event, user comments, metadata, and publisher information. In our experiments, we focus on three core modalities: the video content, the textual description, and the associated news event.

\paragraph{FakeTT.}
FakeTT is an English-language dataset constructed for fake news detection in short videos, collected from mainstream English short-video platforms. It includes a variety of fact-checked news events verified by professional fact-checking organizations. Each sample provides access to the video, the textual event, and related metadata. Consistent with our setup in FakeSV, we primarily use the video, the textual description, and the corresponding news event in our experiments.

Additional statistics can be found in Table~\ref{table3}.

\begin{table*}[t]
  \renewcommand{\arraystretch}{1.2}  
  \caption{The statistical information of both datasets}
  \label{table3}
  \centering
  \begin{tabular}{cccccccc}
    \toprule
    Dataset & Time Range        & Fake & Real & Total & Duration(s) & Language & Platform\\ 
    \midrule
    FakeSV  & 2017/10-2022/02   & 1,810 & 1,814 & 3,624 & 39.88 & Chinese & \makecell{Douyin\\Kuaishou} \\
    FakeTT  & 2019/05-2024/03   & 1,172 & 819   & 1,991 & 47.69 & English & TikTok\\
    \bottomrule
  \end{tabular}
\end{table*}

\subsection{Baselines}
\label{C.2}

To validate the effectiveness of CFGE, we compare it with 13 competitive baselines, which are categorized into three groups.

\subsubsection{VLM based methods}
Utilizing a Vision Language Model for cross-modal reasoning and decision-making. We process each news video using the prompt templates detailed in Appendix ~\ref{A}. For each video, we uniformly sample 8 keyframes, and resize each frame to a resolution of $448 \times 448$ before input. To ensure consistency with the main experiments, we used the 8B version for all open-source VLMs.

\noindent 
$\bullet$ \textbf{GPT-4o-mini:} \cite{r17} A compact and cost-efficient multimodal model developed by OpenAI, GPT-4o-mini supports both text and image inputs, delivering text outputs. It surpasses GPT-3.5 Turbo in academic benchmarks across textual intelligence and multimodal reasoning, making it suitable for resource-constrained applications.

\noindent 
$\bullet$ \textbf{GPT-4.1-mini:} \cite{r18} An enhanced version of OpenAI's small-scale models, GPT-4.1-mini achieves performance comparable to GPT-4o while offering reduced latency and cost. It features a 1 million token context window and excels in tasks requiring long-context understanding and instruction following.

\noindent 
$\bullet$ \textbf{Qwen2.5-VL:} \cite{r19} Developed by Alibaba Group's Qwen team, Qwen2.5-VL is a multimodal vision-language model available in 3B, 7B, 32B and 72B parameter sizes. It introduces advanced features like window attention in the Vision Transformer encoder and dynamic resolution processing, enhancing its capabilities in document parsing and long-video comprehension.

\noindent 
$\bullet$ \textbf{InternVL2.5:} \cite{r20} An advanced Vision Language Model series that builds upon InternVL 2.0, maintaining its core architecture while introducing significant enhancements in training strategies and data quality. InternVL2.5 demonstrates competitive performance across various benchmarks, including multi-discipline reasoning and document understanding.

\noindent 
$\bullet$ \textbf{InternVL2.5-MPO:} \cite{r20} An extension of InternVL2.5 that incorporates Mixed Preference Optimization (MPO) to further enhance multimodal reasoning capabilities.

\subsubsection{Unimodal methods} 
Utilizing only one modality for authenticity determination. 

\noindent 
$\bullet$ \textbf{ViT:} \cite{r13} ViT is a vision transformer model that directly extracts semantic features from image patches using the Transformer architecture. In this work, we sample 8 key frames from each video and use ViT to obtain a 768-dimensional feature vector for each frame. These vectors are subsequently fed into a two-layer MLP to generate the final classification result.

\noindent 
$\bullet$ \textbf{BERT:} \cite{r14} BERT is a pretrained language representation model that captures deep bidirectional semantic information from unlabeled text. We input both the news video description and the news event information into BERT and extract the [CLS] token as the holistic semantic representation. The resulting 768-dimensional feature vector is passed through a two-layer MLP to generate the final classification result.

\subsubsection{multimodal methods}
Utilizing information from multiple modalities for authenticity determination.

\noindent 
$\bullet$ \textbf{TikTec:} \cite{r15} TikTec is a multimodal misinformation detection framework targeting misleading COVID-19 short videos, where deceptive content is jointly expressed across visual, audio, and textual modalities. It leverages video captions to guide the extraction of key visual cues and models the interplay between visual and audio signals. 

\noindent 
$\bullet$ \textbf{FANVM:} \cite{r16} FANVM is a topic-agnostic fake news video detection model that combines adversarial learning and topic modeling. It estimates topic distributions from video titles/descriptions and comments to identify stance inconsistencies, and employs an adversarial neural network to extract topic-independent features. This approach effectively captures cross-modal stance differences and enhances fake news video detection.

\noindent 
$\bullet$ \textbf{SV-FEND:} \cite{r12} SV-FEND is a multimodal detection model built upon the FakeSV dataset, which integrates both news content (including text, audio, keyframes, and video clips) and social context (comments and user profiles). It employs two cross-modal Transformer layers to model the interactions between text and audio, and between text and visual features. A self-attention layer is further used to fuse the content features with social context representations. 

\noindent 
$\bullet$ \textbf{FakingRecipe:} \cite{r9} From the perspective of the creation process of fake news videos, we integrate visual, textual, and audio modalities while simultaneously considering emotional cues present in both text and audio. By modeling the underlying patterns in material selection and editing, the approach aims to uncover deceptive signals embedded in the generation of multimodal content. This enables the detection system to capture anomalies in emotional expression, semantic content, and temporal structure, thereby improving both the accuracy and interpretability of fake news video detection.

\noindent 
$\bullet$ \textbf{CA-FVD:} \cite{r11} Focuses on detecting fake news videos by examining the consistency between different modalities. It utilizes multimodal large language models (MLLMs) to generate pseudo labels that indicate the degree of cross-modal alignment. In addition, it integrates emotional features from both textual and audio inputs to further enhance the model's ability to capture discrepancies and deceptive cues across modalities.

\noindent 
$\bullet$ \textbf{ExMRD:} \cite{r10} ExMRD is an explainable micro-video rumor detection framework that leverages a novel three-step Chain-of-Thought (CoT) inference mechanism—Refining, Retrieving and Reasoning (R\textsuperscript{3}CoT)—to reorganize low-quality content, retrieve domain knowledge, and perform logical reasoning. Instead of fine-tuning large models directly, ExMRD distills the CoT-guided outputs from MLLMs into a lightweight Small Language Reviewer (SLReviewer), ensuring efficient and interpretable predictions. This design enables the model to provide high-quality rationales while maintaining competitive accuracy with reduced computational cost.

\begin{table*}[t]
  \centering
  \begin{tabular}{c|cccc|cccc}
    \hline
    \multicolumn{1}{c|}{\textbf{Dataset}} & \multicolumn{4}{c|}{\textbf{FakeSV}} & \multicolumn{4}{c}{\textbf{FakeTT}} \\
    \hline
     \textbf{Entropy Loss Setting} & ACC & M-F1 & M-P & M-R & ACC & M-F1 & M-P & M-R \\
    \hline
    \multicolumn{1}{c|}{\textit{w/ Entropy Loss}}  & 89.85 & 89.62 & 90.00 & 89.75 & 86.96 & 85.72 & 84.92 & 86.93 \\
    \multicolumn{1}{c|}{\textit{w/o Entropy Loss}} & \textbf{90.22} & \textbf{89.97} & \textbf{90.55} & \textbf{89.64} & \textbf{89.30} & \textbf{87.98} & \textbf{87.80} & \textbf{88.17} \\
    \hline
  \end{tabular}
  \caption{Ablation study on the effect of entropy loss in the PMOE module.}
  \label{table4}
\end{table*}

\subsection{Impact of Entropy Loss}  
\label{C.3}

To verify whether the proposed PMOE module functions as intended, we introduce an additional entropy loss term into the overall loss. Specifically, we extract the probability distribution produced by the four attribution experts and compute the entropy as a regularization term, which is then incorporated into the total loss for joint optimization:
\begin{equation}
\mathcal{L}_{\textit{total}} = \mathcal{L}_{\textit{CE}} + \mathcal{L}_{\textit{PMOE}} + \mathcal{L}_{\textit{ADEC}} + \sum_{i=1}^{4} \mathcal{H}(p_i),
\label{eq21}
\end{equation}
\begin{equation}
\mathcal{H}(p_i) = - p_i \log p_i,
\label{eq22}
\end{equation}
where $p_i$ represents the predicted probability that the news video is attributed to the $i$-th expert.

As shown in Table~\ref{table4}, we observe a decline in performance after introducing the entropy loss. We hypothesize that this is because the entropy term enforces a more deterministic routing of tokens, causing them to concentrate on a single expert. This reduction in expert selection diversity may limit the model’s ability to flexibly assign tokens to the most appropriate expert based on input variations, ultimately leading to degraded performance.

\subsection{Effect of PMOE Placement on Model Performance}
\label{C.4}

We conduct experiments to examine the impact of placing the PMOE module at different layers. As shown in Table~\ref{table5}, placing PMOE at layer 12 alone achieves the highest accuracy, indicating that a single PMOE module is sufficient for effective manipulation reasoning. In contrast, placements at layers 8 and 12 or 12 and 16 slightly reduce performance, suggesting that misaligned injection timing may weaken the interaction between manipulation cues and contextual representations.

\begin{table}[t]
  \centering
  \begin{tabular}{c|cccc}
    \hline
    \textbf{Layers} & \textbf{ACC} & \textbf{M-F1} & \textbf{M-P} & \textbf{M-R} \\
    \hline
     8 \& 12   & 88.96 & 87.63 & 87.37 & 87.92 \\
     12 \& 16  & 88.96 & 87.63 & 87.37 & 87.92  \\
     12 only   & \textbf{89.30} & \textbf{87.98} & \textbf{87.80} & \textbf{88.17} \\
    \hline
  \end{tabular}
  \caption{Experiments of PMOE placement at different LLM layers on the FakeTT dataset.}
  \label{table5}
\end{table}

\subsection{Quantitative analysis of MoE routing results}
\label{C.5}

We further analyze the quantitative routing behavior of our Progressive MoE (PMOE) by examining the Detection MoE and Attribution MoE separately. 

\subsubsection{Detection MoE.} Its gating probabilities indicate the likelihood of a news video being classified as real or fake (e.g., [0.9, 0.1]). We analyzed all 299 samples in the FakeTT test set, and the results demonstrate strong consistency with the final test performance (ACC: 89.30, M-F1: 87.98, M-P: 87.80, M-R: 88.17). The detailed confusion matrix is shown in Table~\ref{table6}.

\begin{table}[t]
  \centering
  \begin{tabular}{c|cccc}
    \hline
    \textbf{Category} & \textbf{TP} & \textbf{FN} & \textbf{FP} & \textbf{TN} \\
    \hline
    Prediction & 183 & 17 & 15 & 84 \\
    \hline
  \end{tabular}
  \caption{Confusion matrix of Detection MoE on FakeTT test set.}
  \label{table6}
\end{table}

\subsubsection{Attribution MoE.} Its gating probabilities indicate the predicted probabilities for four manipulation types: \textit{real}, \textit{visual manipulation}, \textit{textual manipulation}, and \textit{fully manipulated}. Due to the absence of fine-grained ground-truth labels, we manually annotated 100 samples from the FakeTT test set (with class counts of 56, 17, 17, and 10, respectively). The results are summarized in Table~\ref{table7}, showing strong consistency with our manual annotations (ACC: 82.00, M-F1: 76.19, M-P: 74.62, M-R: 78.63).

\begin{table*}[t]
  \centering
  \begin{tabular}{c|cccc}
    \hline
     & \textbf{Real} & \textbf{Visual manipulation} & \textbf{Textual manipulation} & \textbf{Fully manipulated} \\
    \hline
    Real                & 48 & 5  & 2  & 1 \\
    Visual manipulation & 1  & 13 & 1  & 2 \\
    Textual manipulation& 0  & 2  & 14 & 1 \\
    Fully manipulated   & 0  & 2  & 1  & 7 \\
    \hline
  \end{tabular}
  \caption{Confusion matrix of Attribution MoE on 100 manually annotated FakeTT samples.}
  \label{table7}
\end{table*}

\begin{figure*}[t]
 \centering
  \begin{subfigure}[b]{\textwidth}
    \includegraphics[width=\textwidth]{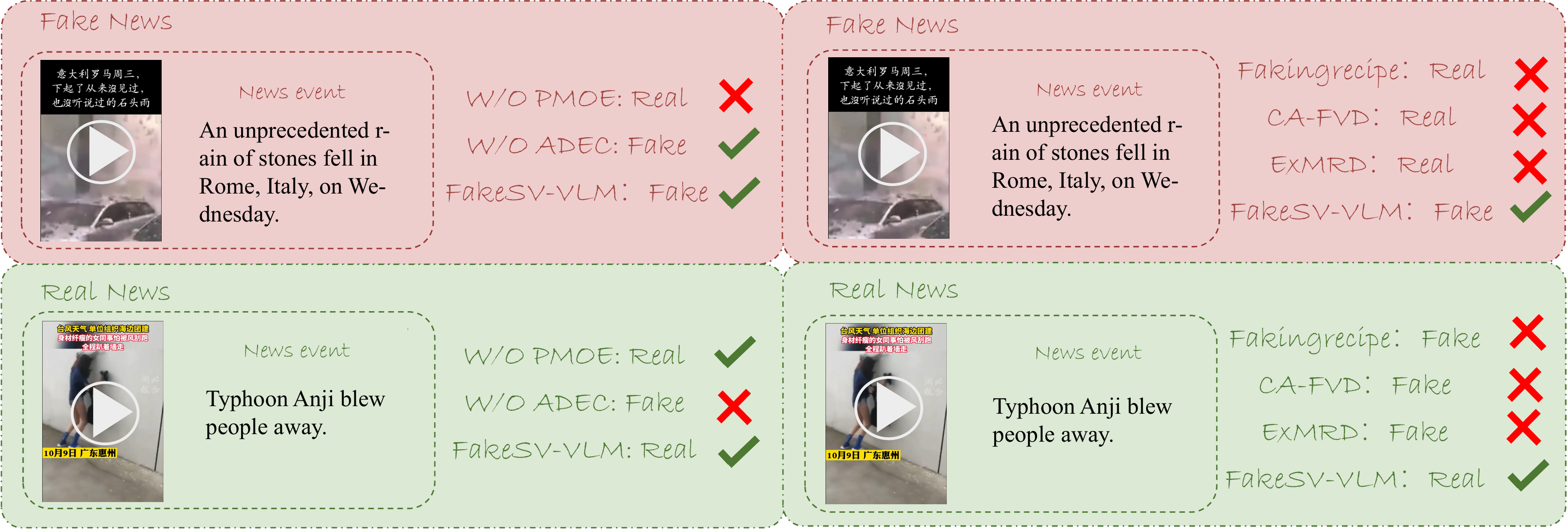}
    \caption{Two successful cases from the FakeSV dataset.}
    \label{Fig5a}
  \end{subfigure}

  \begin{subfigure}[b]{\textwidth}
    \includegraphics[width=\textwidth]{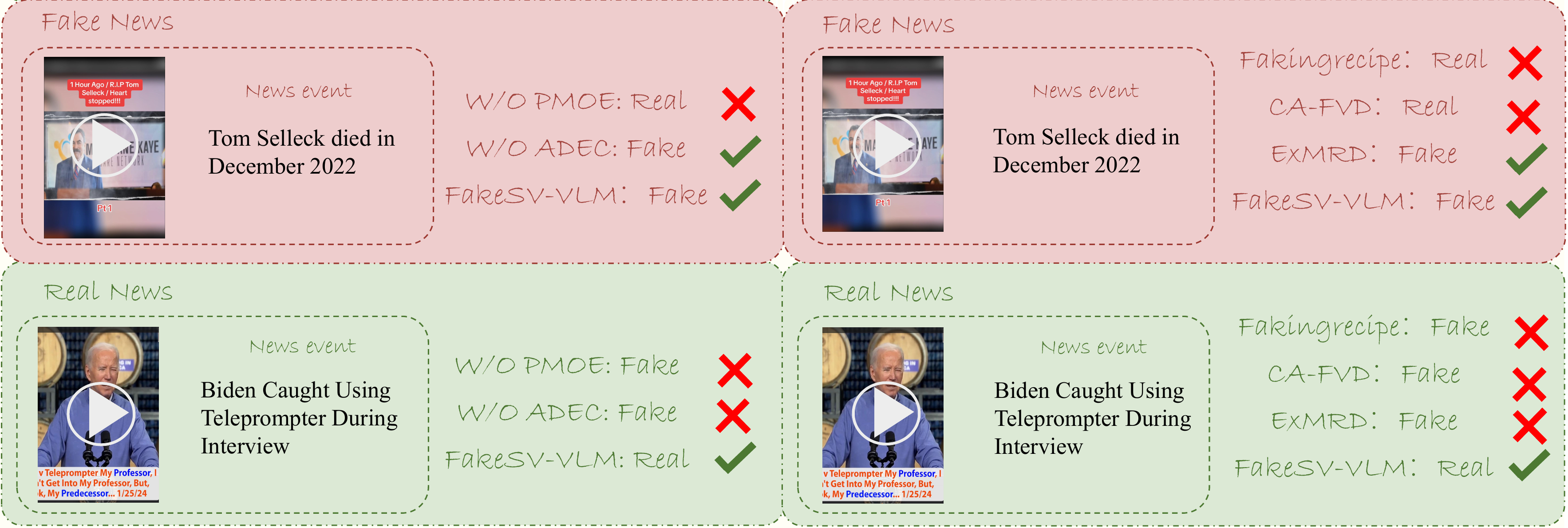}
    \caption{Two successful cases from the FakeTT dataset.}
    \label{Fig5b}
  \end{subfigure}
  
  \caption{Four successful cases of fake news videos selected from the FakeSV and FakeTT datasets.}
  \label{Fig5}
\end{figure*}

\begin{figure*}[ht!]
 \centering
  \begin{subfigure}[b]{\textwidth}
    \includegraphics[width=\textwidth]{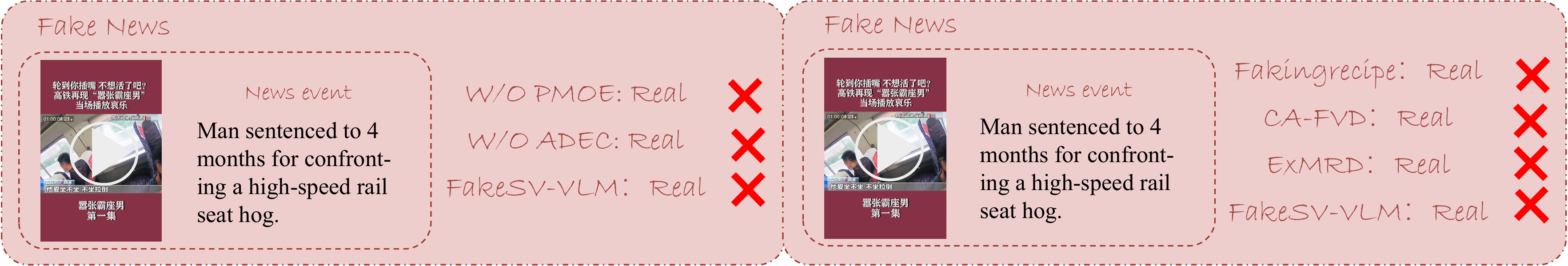}
    \caption{A failure cases from the FakeSV dataset.}
    \label{Fig6a}
  \end{subfigure}

  \begin{subfigure}[b]{\textwidth}
    \includegraphics[width=\textwidth]{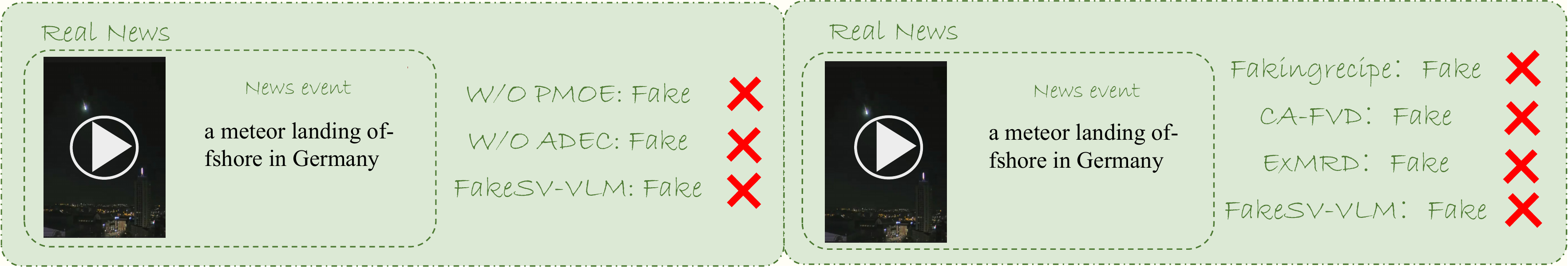}
    \caption{A failure cases from the FakeTT dataset.}
    \label{Fig6b}
  \end{subfigure}
  
  \caption{Two failure cases of fake news videos selected from the FakeSV and FakeTT datasets.}
  \label{Fig6}
\end{figure*}

\section{Other Case Studies}
\label{D}

\subsection{Successful Cases}
\label{D1}

As illustrated in Figure~\ref{Fig5}, we present two successful cases from the FakeSV dataset in ~\ref{Fig5a} and two cases from the FakeTT dataset in ~\ref{Fig5b} to demonstrate the effectiveness of our proposed FakeSV-VLM framework. For consistency, we translate the news video content in the FakeSV dataset into Chinese.

The first case is titled \textit{"An unprecedented rain of stones fell in Rome, Italy, on Wednesday"}, which falsely describes a fabricated weather phenomenon. Models such as FakingRecipe, CA-FVD and ExMRD misclassify the video as real news. Without the PMOE module, the model fails to detect the manipulation. However, after incorporating PMOE, the model successfully identifies the abnormality by leveraging expert reasoning over the overall authenticity and attribution type, ultimately leading to the correct classification as fake news.

The seconde case is titled \textit{"Typhoon Anji blew people away"}, which presents a real news event but is exaggerated through dramatic captioning. While FakingRecipe, CA-FVD and ExMRD misclassify the video as fake news, our method produces the correct prediction. Without ADEC, the model fails to resolve the inconsistency. After incorporating the ADEC module, the model effectively verifies the alignment between the visual content and the claimed event semantics, enabling it to correctly identify the video as real news.

The third case is titled \textit{"Tom Selleck died in December 2022"}, which falsely claims the death of the well-known actor. FakingRecipe and CA-FVD fail to recognize the manipulation. With the integration of the PMOE module, the model first evaluates the overall authenticity of the video through the Detection MoE, and then further infers the potential type of manipulation via the Attribution MoE, enabling it to correctly identify the video as fake news.

The fourth case involves the event \textit{"Biden Caught Using Teleprompter During Interview"}. Although the footage in the video is real, the claim is taken out of context to suggest deception. FakingRecipe, CA-FVD and ExMRD misclassify this video as fake news. Building upon the hierarchical reasoning of overall authenticity and manipulation types provided by the PMOE module, the ADEC module further captures fine-grained semantic inconsistencies between the narrative content and the visual-textual evidence. This collaboration enables the model to more accurately identify context-based manipulations and make correct predictions.

\subsection{Failure Cases}
\label{D2}

As illustrated in Figure~\ref{Fig6}, we present a failure case from the FakeSV dataset in ~\ref{Fig6a} and a from the FakeTT dataset in ~\ref{Fig6b} to highlight the limitations of our proposed FakeSV-VLM framework in handling challenging scenarios.

The first case is titled \textit{"Man sentenced to 4 months for confronting a high-speed rail seat hog"}, which is in fact a fabricated news event. Nevertheless, due to the presence of real-world footage and a highly plausible narrative structure, the video appears highly credible in form. As a result, multiple detection models—including our proposed FakeSV-VLM—misclassify the video as real news. This outcome highlights the potential risk of misjudgment when models encounter carefully manipulated content that incorporates authentic visual elements.

The second case is titled \textit{"A meteor landing offshore in Germany"}, which describes a real event captured from a public livestream. Despite being factually correct, the video was misclassified as fake by several models. We suspect that the unusual visual features and lack of explicit semantic grounding in the textual description hindered the alignment process. Such mismatch makes it difficult for the model to establish a clear cross-modal correspondence. This suggests that when natural phenomena appear visually ambiguous or rare, the model may overfit to visual irregularity signals and fail to correctly align them with benign event semantics.

\end{document}